\documentclass{elsart}
\journal{Physica D}

\usepackage{amssymb}
\usepackage{amsmath}
\usepackage{graphicx}
\usepackage{psfrag}

\renewcommand{\atop}[2]{\genfrac{}{}{0pt}{}{#1}{#2}}

\newcommand{\rmi}{\mathrm{i}}
\newcommand{\rme}{\mathrm{e}}
\newcommand{\eref}[1]{(\ref{#1})}
\def \mc{\mathcal}
\def \tw{\textwidth}

\def \rm{\textrm}

\begin{document}


\begin{frontmatter}
\title{Stability, mobility and power currents in a two-dimensional 
model for waveguide arrays with nonlinear coupling}
\author{Michael \"{O}ster},
\ead{micos@ifm.liu.se}
\author{Magnus Johansson\corauthref{cor}}
\corauth[cor]{Corresponding author.}
\ead{mjn@ifm.liu.se}
\ead[url]{http://people.ifm.liu.se/majoh}
\address{Department of Physics, Chemistry and Biology (IFM),
Link\"{o}ping University, SE-581 83 Link\"{o}ping, Sweden}

\begin{abstract}
A two-dimensional nonlinear Schr\"{o}dinger lattice with nonlinear 
coupling, modelling a square array of weakly coupled linear optical waveguides 
embedded in a nonlinear Kerr material, is studied. We find that despite a 
vanishing 
energy difference (Peierls-Nabarro barrier) of fundamental stationary modes 
the mobility of localized excitations is very poor. This is attributed to 
a large separation in parameter space of the bifurcation points of the 
involved stationary modes. At these points the stability of the fundamental 
modes is changed and an asymmetric intermediate solution appears that 
connects the points. The control of the power flow across the array when 
excited with plane waves is also addressed and shown to exhibit great 
flexibility that may lead to applications for power-coupling devices. 
In certain parameter regimes, the direction of a stable propagating 
plane-wave current  is shown to be continuously 
tunable by amplitude variation (with fixed phase gradient). 
More exotic effects of the nonlinear coupling terms like compact discrete 
breathers and vortices, and stationary complex modes with non-trivial phase 
relations are also briefly discussed. Regimes of dynamical linear stability 
are found for all these types of solutions. 
\end{abstract}

\begin{keyword}
Nonlinear coupling, Peierls-Nabarro potential, mobility, inversion of 
stability, modulational instability, power currents.
\PACS 42.65.Wi \sep 42.82.Et \sep 63.20.Pw \sep 05.60.-k
\end{keyword}

\end{frontmatter}

\section{Introduction}
\label{sec:introduction}
The use, in optics, of materials with nonlinear responses to external fields 
has over the past years provided a wealth of applications and prospects for 
all-optical communication, where light manipulates light 
itself~\cite{Kivshar.03}. Especially, nonlinear photonic structures endowed 
with a spatially periodic refractive index provide new means to control the 
flow of light in the form of localized wave packets, not possible in bulk 
media~\cite{Christodoulides.03}. The experimental observation of discrete 
localization in one and two dimensions, in arrays of coupled waveguides 
experiencing a cubic (Kerr) 
nonlinearity~\cite{Eisenberg.98,Morandotti.99,Morandotti.01,Pertsch.04} or a 
saturable nonlinearity through the photovoltaic effect in a photorefractive 
medium~\cite{Chen.05} as well as in optically induced nonlinear photonic 
lattices~\cite{Fleischer.03,Fleischer.03b,Neshev.04,Fleischer.04}, shows 
significant progress in this direction. The theoretical foundation and 
predictions of these phenomena can, to good approximation, be made by models 
of the discrete nonlinear Schr\"{o}dinger (DNLS) 
type~\cite{Pertsch.04,Jensen.82,Christodoulides.88,Stepic.04,Efremidis.02,Mingaleev.01} 
derived within coupled-mode theory~\cite{Yariv.73} or the tight-binding 
approximation using Wannier function expansion~\cite{Alfimov.02}. The latter 
technique is also applicable for models of Bose-Einstein condensates trapped 
in deep optical lattices. In these contexts the issue of mobility of 
localized packets of energy is highly interesting as it provides insights as 
how to control optical signals, e.g., for applications in multiport switching.

As is well known~\cite{Scott.99}, travelling localized solutions, or 
solitons, quite generally exist in continuum systems as a result of the 
competition between nonlinearity and dispersion. When the continuous symmetry 
is broken, as is the case when considering a nonlinear discrete system, the 
generic solution is instead a non-moving time-periodic one, called a discrete 
breather (DB) or intrinsic localized mode (see, e.g.,~\cite{Campbell.04} for 
a review). The presence of a lattice leads to pinning of localized solutions, 
as this underlying structure effectively introduces a periodic potential for 
travelling excitations due to the difference in energy between stationary 
solutions localized on a lattice site and solutions localized between lattice 
sites~\cite{Claude.93,Kivshar.93}. This energy difference is referred to as a 
Peierls-Nabarro (PN) energy barrier and is the reason that narrow travelling 
excitations get trapped in a lattice. Mobility of a highly localized 
stationary mode can be induced if an appropriate perturbation is added to 
overcome the potential barrier, but usually some energy is lost to radiation 
and the result is trapping~\cite{Chen.96,Bang.96}. However, excitations that 
are broad, in comparison with the lattice spacing, can still propagate as the 
effects of the PN-barrier in this case is small and the dynamics is well 
approximated by the continuum limit of the discrete equation. This is 
provided when the continuum equation has stable localized travelling 
solutions, which, e.g., is the case for the 1D cubic nonlinear 
Schr\"{o}dinger equation but not for the 2D (cubic) 
equation~\cite{Rasmussen.86}.

Notable exceptions from the pinning behaviour are integrable lattice 
equations, i.e., equations possessing an infinite number of conserved 
quantities, like the Ablowitz-Ladik (AL) model which is a discretization of 
the nonlinear Schr\"{o}dinger equation retaining this 
property~\cite{Ablowitz.75}. These models have analytic exponentially 
localized travelling solutions and completely lack PN-barrier, which accounts 
for the excellent mobility. Unfortunately, in contrast to their many 
interesting mathematical properties, they seem to have very limited direct 
applications for real systems. Thus other routes to mobility in discrete 
systems are sought, although these models are conveniently used as the 
starting point of perturbative approaches. In~\cite{Gomez.04} it is 
numerically shown how a mobile excitation develops a resonant tail when the 
governing equation deviates from the 1D integrable AL-model. Moreover, the 
amplitude of the tail grows when the equation is further away from the 
integrable limit, or alternatively, as the PN-barrier increases from zero. 
It is inferred that the effect of pinning is overcome by a periodic exchange 
of energy between the localized core and the oscillating tail as the 
excitation moves through the lattice and the periodic PN-potential. This 
result can also be extended to a 2D setting~\cite{Gomez.06}. However, there 
is a limit as to how far the mobile solutions can be continued away from the 
integrable limit, with broad solutions being more persistent as can be 
expected~\cite{Gomez.04}. Hence, we can expect that mobility of localized 
modes is possible when the PN-barrier is small, and the ability to tailor the 
size of the PN-barrier may allow for a control of the mobility of narrow 
solutions in a lattice.

Recent research has focused on ways to minimize the PN-barrier, which in 
optics applications means moving beyond models based on the cubic DNLS 
equation where the PN-barrier is an increasing function of the power of the 
stationary modes~\cite{Morandotti.99}. Similar to the problem of moving kinks 
in Klein-Gordon lattices~\cite{Savin.00} the introduction of competing 
nonlinearities has proven fruitful. In DNLS equations with saturable 
nonlinearity the vanishing of the PN-barrier at certain points and a 
subsequent enhanced mobility has been demonstrated in both 
1D~\cite{Hadzievski.04} and 2D~\cite{Vicencio.06}, and the same apply for a 
1D DNLS model with nonlinear interactions~\cite{Oster.03}. As these models 
are derived from real physical settings they are interesting from an 
application point of view. Note also that there is a large class of DNLS 
models~\cite{Dmitriev.06} with an absence of PN-barrier (not only zero at 
isolated points), different from the more physical models, as the former are 
special discretizations of the nonlinear Schr\"{o}dinger model that in some 
sense preserve a translational invariant~\cite{Pelinovsky.06,Kevrekidis.06}. 
Amongst them is the AL-lattice and other integrable equations.

A problem related to the vanishing of the PN-barrier at specific points is 
that of stability inversion between stationary solutions centred on a lattice 
point and between lattice points, respectively, which can be understood from 
the energy minimum principle. The solution with lowest energy at given power 
will be stable, and the change of sign of the energy difference is also 
related to an exchange of stability of the lowest energy modes. This can be 
observed for the above 
models~\cite{Hadzievski.04,Vicencio.06,Oster.03,Maluckov.05,Maluckov.06}, as 
well as for the DNLS model with competing cubic-quintic 
nonlinearity~\cite{Carretero.06} and more general lattice 
models~\cite{Cretegny.98,Aubry.06}. In connection with the stability 
inversion there also exists an intermediate asymmetric solution interpolating 
between the stationary solutions centred on and in-between sites and 
bifurcating with these at the points where they change their stability 
properties~\cite{Vicencio.06,Oster.03,Carretero.06,Cretegny.98,Aubry.06}. For 
the 1D saturable DNLS model this solution can even be found analytically near 
the first zero of the PN-barrier~\cite{Khare.05}. From this viewpoint, the 
mobility of a highly localized mode may be thought of as a transformation 
between stationary on- and off-site solutions, via the intermediate solution. 
Hence a more accurate definition of the PN-barrier would need to include also 
the intermediate solution as this may be the mode with the highest energy. 
Therefore, we will distinguish between the \emph{energy difference} between 
two fundamental modes and the \emph{PN-barrier} relating to the energy of all 
involved stationary modes. Also, in 2D, one would in principle need to 
consider modes corresponding to different directions of possible 
propagation~\cite{Vicencio.06}.

Though localized modes and their mobility are interesting for beam steering 
in waveguide arrays, transport of energy in the transversal direction (from 
waveguide to waveguide) can be achieved also by a collective excitation of 
the waveguides. With a plane wave propagating in the array, the same power 
will be transported along all waveguides, but there will also be a transfer 
of power between the waveguides depending on the wave number (phase gradient) 
of the plane wave. In~\cite{Oster.05} it is demonstrated that the direction 
and magnitude of the transversal flow of power in a 1D array with nonlinear 
coupling are quite versatile under the variation of the amplitude and wave 
number of the plane waves. Here the discussion is extended to 2D and will 
show an even greater flexibility.

The outline of the Paper is as follows: In Sec.~\ref{sec:model} we present a 
model for optical waveguides coupled in two spatial dimensions and embedded 
in a nonlinear Kerr material, discuss its relevance and introduce some of the 
important properties of the model. We should stress that our model does 
\emph{not} attempt to describe configurations where the waveguides themselves 
are nonlinear, such as those presently 
experimentally realized in, e.g., 
\cite{Eisenberg.98,Morandotti.99,Morandotti.01,Pertsch.04,Chen.05,Fleischer.03,Fleischer.03b,Neshev.04,Fleischer.04}, but rather correspond to the original 
proposal of Jensen \cite{Jensen.82} with the nonlinearity residing solely 
in the \emph{embedding} material. The stability of the fundamental localized 
modes 
are investigated in Sec.~\ref{sec:stationary} and their relation to mobility 
of localized excitations is discussed in Sec.~\ref{sec:mobility}. In 
Sec.~\ref{sec:compact} some special solutions, like compact and complex 
modes, arising from a balance of the coupling terms in the model are 
presented. Finally, in Sec.~\ref{sec:planewave} the transport properties of 
plane waves are analyzed before we conclude in Sec.~\ref{sec:conclusion}.
We also include in the end of Sec.~\ref{sec:conclusion} a brief discussion 
about some 
recent related works, which appeared in press 
after the original submission of this manuscript. 

\section{Model}
\label{sec:model}
The model we consider is a direct extension to two spatial dimensions of a 
one-dimensional model for the propagation of the electric field amplitudes in 
coupled optical waveguides embedded in a nonlinear material~\cite{Oster.03}. 
The waveguides are identical and regularly spaced in the two spatial 
dimensions, thus forming a square lattice. Coupling between the sites of the 
lattice is mediated by an evanescent field overlap of the modes (single-mode 
propagation) in the respective waveguides and it is assumed that the modes 
decay sufficiently fast outside the waveguides to motivate only 
nearest-neighbour coupling. Further, the waveguides themselves are assumed to 
be constructed from a \emph{linear} material and 
\emph{surrounded by a nonlinear Kerr material}. This will tend to strengthen 
the effects of nonlinear coupling with respect to on-site nonlinear effects 
when compared to a system where also the waveguides are nonlinear. This 
motivates an extension of the generally employed cubic on-site DNLS model to 
an equation that in two dimensions will read
\begin{equation}
\begin{split}
\label{eq:xdnls}
   \rmi\dfrac{d\Psi_{n,m}}{dz} = &
   \phantom{0}Q_{1}\Psi_{n,m}
   + Q_{2}\Delta\Psi_{n,m}
   + 2Q_{3}\Psi_{n,m}|\Psi_{n,m}|^{2} \\
   & + 2Q_{4}\big[2\Psi_{n,m}\Delta(|\Psi_{n,m}|^{2})
      + \Psi_{n,m}^{*}\Delta(\Psi_{n,m}^{2}) \big] \\
   & + 2Q_{5}\big[2|\Psi_{n,m}|^{2}\Delta\Psi_{n,m}
      + \Psi_{n,m}^{2}\Delta\Psi_{n,m}^{*} \\
   & \phantom{+2Q_{5}[0}
      + \Delta(\Psi_{n,m}|\Psi_{n,m}|^{2}) \big],
\end{split}
\end{equation}
where $\Psi_{n,m}$ is the complex amplitude of the electric field, 
$z$ measures the propagation along the waveguide array and the coupling 
parameters $Q_{1}$--$Q_{5}$ are given by overlap integrals of the modes. The 
operator $\Delta$ is defined by 
$\Delta\Psi_{n,m}=\Psi_{n-1,m}+\Psi_{n+1,m}+\Psi_{n,m-1}+\Psi_{n,m+1}$. 
(Note that the operator $\Delta -4$ is the standard 2D discrete Laplacian.)

The model~\eref{eq:xdnls} is also relevant in other contexts, e.g., as a 
nonlinear tight-binding approximation, with $Q_{4}$ argued to be negligible, 
for the time-evolution of Bose-Einstein condensates (BEC) in a deep periodic 
optical potential~\cite{Smerzi.03,Smerzi.03b,Menotti.03}, or for 
$Q_{3}/2=Q_{4}=Q_{5}$ as a rotating-wave approximation to a Fermi-Pasta-Ulam 
chain~\cite{Claude.93}. In both these one-dimensional models the extension to 
two spatial dimensions is straightforward. One may also in a straightforward 
way consider extensions to other lattice geometries, such as 
e.g.\ triangular/hexagonal patterns (which in some sense could be considered 
as the most natural packing of waveguides in two dimensions). In this work, 
we chose to consider only the square lattice which is the simplest 2D 
structure to analyze from a modelling point of view, and also for 
comparison with previously published work in the area which mainly dealt
with square-lattice configurations. We expect that most of the qualitative 
results obtained here will persist also for different lattice geometries, 
although a more detailed discussion on these issues is beyond the scope 
of the present work. 

\subsection{Parameter estimates}
\label{sec:parameters}
For general waveguide arrays $Q_{2}>0$ and $Q_{j}$, $j=3,4,5$, have the same 
sign as the Kerr index of the nonlinear material. Estimates of the relative 
strengths of the parameters show that the nonlinear coupling terms, $Q_{4}$ 
and $Q_{5}$, can be up to the same order of magnitude as the on-site 
nonlinearity $Q_{3}$ for an array of waveguides constructed from AlGaAs with 
sizes in the tenth of micrometer regime and operated with a laser in the 
infrared ($\lambda\approx 1.5 \mu m$). Details of the derivation of the 1D 
model can be found in~\cite{Oster.03,Oster.05b} and a calculation of the 
coupling parameters for a slab waveguide array can be found 
in Appendix A of~\cite{Oster.05b} (see in particular 
Fig.\ A2 in~\cite{Oster.05b} for an explicit illustration of realistic values 
for relative coupling strengths). When extending the model to two dimensions 
there is also the possibility of coupling in the diagonal direction between 
next-nearest neighbours. However, for experimentally relevant sizes, as given 
above, of an array of square waveguides the linear diagonal coupling can be 
estimated to be an order of magnitude smaller than the linear direct 
coupling (see Appendix A2 in \cite{Oster.05b}). Also, when operated in the 
power regime where the most interesting phenomena occur, the nonlinear 
coupling and on-site terms will be of the same order as the linear coupling, 
i.e. $|Q_{2}|\sim |Q_{j}|\sup\{|\Psi_{n,m}|^{2}\}$, $j=3,4,5$. This can be 
shown to correspond to powers of a few kW~\cite{Oster.05b}, well within what 
is experimentally available. However, although these estimates of the 
parameter values are of relevance when considering experimental verifications 
and applications of the model they will not be taken as a restriction in the 
present paper, since we are also interested in the general effects of 
nonlinear coupling.

Further, the parameter $Q_{1}$ can be made to vanish by the transformation 
$\Psi_{n,m}(z)\mapsto\Psi_{n,m}(z)\rme^{-\rmi Q_{1}z}$ and has no significant 
physical role. Note also that the staggering transformation 
$\Psi_{n,m}(z)\mapsto(-1)^{n+m}\Psi_{n,m}(z)$ is equivalent to changing the 
sign of the parameters $Q_{2}$ and $Q_{5}$, thus reducing the part of 
parameter space that needs to be investigated to get a complete picture of 
the equation. Further, by rescaling the amplitudes and the time-like 
variable, two additional parameters can be fixed. Here we will fix 
$Q_{2}=0.2$ and $Q_{3}=0.5$, while restricting $Q_{4}\geq 0$ without loss of 
generality.

\subsection{Conserved quantities}
\label{sec:conserved}
From Noether's theorem we know that any infinitesimal transformation leaving 
the action integral for an equation invariant leads to a quantity that is 
conserved under the evolution governed by that equation~\cite{Goldstein.80}. 
Essentially this means that any continuous symmetry of Eq.~\eref{eq:xdnls} 
corresponds to a conserved quantity. In particular, the invariance under 
translations in $z$ will lead to conservation of the Hamiltonian 
$\mc{H}=\sum_{n,m}\mc{H}_{n,m}$, with
\begin{equation}
\begin{split}
\label{eq:H}
   \mc{H}_{n,m} = \Big\{ & Q_{2}(\Psi_{n,m}\Psi_{n+1,m}^{*}
      +\Psi_{n,m}\Psi_{n,m+1}^{*})
   + \frac{Q_{3}}{2}|\Psi_{n,m}|^{4} \\
   & + Q_{4}\big[ 2|\Psi_{n,m}|^{2}(|\Psi_{n+1,m}|^{2}+|\Psi_{n,m+1}|^{2}) \\
   & \phantom{+Q_{4}0}
      + \Psi_{n,m}^{2}(\Psi_{n+1,m}^{*\,2}+\Psi_{n,m+1}^{*\,2}) \big]  \\
   & + 2Q_5\big[\Psi_{n,m}\Psi_{n+1,m}(\Psi_{n,m}^{*\,2}+\Psi_{n+1,m}^{*\,2}) 
\\
   & \phantom{+ 2Q_50}
      +\Psi_{n,m}\Psi_{n,m+1}(\Psi_{n,m}^{*\,2}+\Psi_{n,m+1}^{*\,2})\big]
      \Big\} + c.c. \,.
\end{split}
\end{equation}
The evolution equation~\eref{eq:xdnls}, and its complex conjugate, can be 
obtained from the Hamilton equations of motion using the canonical variables 
$\Psi_{n,m}$ and $\rmi\Psi_{n,m}^{*}$,
\begin{subequations}
\label{eq:Heq}
\begin{align}
\label{eq:Heq1}
   \rmi\dfrac{d\Psi_{n,m}}{dz} =
      \dfrac{\partial\mc{H}}{\partial\Psi_{n,m}^{*}}, \\
\label{eq:Heq2}
   -\rmi\dfrac{d\Psi_{n,m}^{*}}{dz} =
      \dfrac{\partial\mc{H}}{\partial\Psi_{n,m}}.
\end{align}
\end{subequations}
Further, the invariance of Eq.~\eref{eq:xdnls} under global phase rotations, 
i.e., $\Psi_{n,m}\mapsto\Psi_{n,m}\rme^{\rmi\alpha}$, $\alpha\in\mathbb{R}$, 
corresponds to the conservation of the norm
\begin{equation}
\label{eq:N}
   \mc{N} = \sum_{n,m}\mc{N}_{n,m} = \sum_{n,m}|\Psi_{n,m}|^{2} \,.
\end{equation}
Although the Hamiltonian $\mc{H}$ for all practical purposes constitutes an 
energy functional it has no direct connection to any physical energy for the 
array of waveguides. But from a mathematical viewpoint it can be treated as 
the energy of the system. In particular, for the cubic DNLS model it holds 
that a bounded extremum (a maximum in the present formulation since 
$Q_{2},Q_{3}>0$ and hence $\mc{H}$ should be regarded as negative energy) of 
$\mc{H}$ for given norm $\mc{N}$ is Lyapunov stable and may, with right, be 
called the ground state of the system~\cite{Weinstein.99}. This ground state 
is conjectured to be unique up to a global phase factor, which is consistent 
with the absence of stability inversion and the non-vanishing PN-barrier in 
the cubic on-site DNLS model. The norm $\mc{N}$, however, has a direct 
physical interpretation as the total power carried along the waveguides, 
provided that the modes are appropriately normalized. In the context of BEC, 
the conservation of norm corresponds to boson number conservation.

The conservation of Hamiltonian and norm may also conveniently be expressed 
in terms of discrete continuity equations. For this purpose it is instructive 
to make a change to action-angle variables 
$\Psi_{n,m}=\sqrt{\mc{N}_{n,m}}\,\rme^{-\rmi\theta_{n,m}}$, where 
$\mc{N}_{n,m}$ and $\theta_{n,m}$ are real canonical variables. 
The corresponding Hamilton equations of motions are
\begin{subequations}
\label{eq:Heq_aa}
\begin{align}
\label{eq:Heq_theta}
   \dfrac{d\theta_{n,m}}{dz} & = \dfrac{\partial\mc{H}}{\partial\mc{N}_{n,m}}, 
\\
\label{eq:Heq_N}
   -\dfrac{d\mc{N}_{n,m}}{dz} & = \dfrac{\partial\mc{H}}{\partial\theta_{n,m}}.
\end{align}
\end{subequations}
Now, with $\mc{H}_{n,m}$ from Eq.~\eref{eq:H} and by introducing the phase 
difference between neighbouring sites in the two lattice directions as 
$\varphi_{n,m}=\theta_{n,m}-\theta_{n-1,m}$ and 
$\phi_{n,m}=\theta_{n,m}-\theta_{n,m-1}$ Eq.~\eref{eq:Heq_N} for the norm 
density can be written as
\begin{equation}
\label{eq:Neq}
   \dfrac{d\mc{N}_{n,m}}{dz} + \mc{J}_{n,m}^{(n)} - \mc{J}_{n-1,m}^{(n)}
   + \mc{J}_{n,m}^{(m)} - \mc{J}_{n,m-1}^{(m)} = 0.
\end{equation}
The norm current density in the $\hat{\mathbf{n}}$ direction is given by
\begin{equation}
\label{eq:JN}
\begin{split}
  \mc{J}_{n,m}^{(n)} = & -\dfrac{\partial\mc{H}_{n,m}}{\partial\varphi_{n+1,m}}
   = 2\sqrt{\mc{N}_{n,m}\mc{N}_{n+1,m}}\sin\varphi_{n+1,m} \\
   & \times\big[Q_{2}
      +4Q_{4}\sqrt{\mc{N}_{n,m}\mc{N}_{n+1,m}}\cos\varphi_{n+1,m} \\
   & \phantom{\times\big[} +2Q_{5}(\mc{N}_{n,m}+\mc{N}_{n+1,m})\big].
\end{split}
\end{equation}
A similar expression holds for 
$\mc{J}_{n,m}^{(m)}=-\partial\mc{H}_{n,m}/\partial\phi_{n,m+1}$ in the 
$\hat{\mathbf{m}}$ direction where the roles of $n$ and $m$ are interchanged. 
Since the norm density $\mc{N}_{n,m}$ is the power on site $(n,m)$, the 
continuity equation expresses the obvious fact that a change of the power is 
related to a power transfer to neighbouring sites. Hence the physical 
interpretation of $\mc{J}_{n,m}^{(n)}$ is the flow of power from site $(n,m)$ 
to site $(n+1,m)$. An equation on the form~\eref{eq:Neq} can also be derived 
for the Hamiltonian density $\mc{H}_{n,m}$, with a Hamiltonian flux density 
$\mc{I}_{n,m}^{(n)}$ and $\mc{I}_{n,m}^{(m)}$. For the present purposes it is 
sufficient to know that for a stationary monochromatic solution, i.e., 
$\mc{N}_{n,m}(z)=\mc{N}_{n,m}$ and $\theta_{n,m}(z)=\theta_{n,m}+\Lambda z$, 
the relation to the norm current density is 
$\mc{I}_{n,m}^{(n)}=\Lambda\mc{J}_{n,m}^{(n)}$. Our discussions on the flow 
of energy in the system (Sec.~\ref{sec:planewave}) will thus be made using 
only the norm current density.

\section{Stationary solutions}
\label{sec:stationary}
Initially we will turn our attention towards localized solutions or, more 
specifically, to time-periodic localized modes also known as discrete 
breathers (DBs) or intrinsic localized modes. The existence of DBs has been 
rigorously proven for Hamiltonian lattice systems of arbitrary dimension 
subject to conditions of anharmonicity and non-resonance with linear 
phonons~\cite{MacKay.94}, and extended to more general systems 
in~\cite{Sepulchre.97}. In the context of coupled waveguides the variable $z$ 
will play the role of time. Due to the global phase invariance of the 
DNLS-type equations the time-periodicity leads to a significant 
simplification of the problem of finding localized solutions. For 
monochromatic solutions a simple gauge transformation to a rotating frame of 
reference will render these solutions stationary, i.e., 
taking $\Psi_{n,m}(z)=\psi_{n,m}\rme^{-\rmi(\Lambda+Q_{1})z}$ reduces 
Eq.~\eref{eq:xdnls} to
\begin{equation}
\begin{split}
\label{eq:xdnlsstat}
   & -\Lambda\psi_{n,m}
   + Q_{2}\Delta\psi_{n,m}
   + 2Q_{3}\psi_{n,m}|\psi_{n,m}|^{2} \\
   & + 2Q_{4}\big[2\psi_{n,m}\Delta(|\psi_{n,m}|^{2})
      + \psi_{n,m}^{*}\Delta(\psi_{n,m}^{2}) \big] \\
   & + 2Q_{5}\big[2|\psi_{n,m}|^{2}\Delta\psi_{n,m}
      + \psi_{n,m}^{2}\Delta\psi_{n,m}^{*} \\
      & + \Delta(\psi_{n,m}|\psi_{n,m}|^{2}) \big] = 0.
\end{split}
\end{equation}
The idea of the existence proofs, and also the foundation for efficient 
numerical techniques to calculate DBs~\cite{Eilbeck.84,Marin.96}, is 
continuation from the anti-continuous limit, $Q_{j}=0$, $j=2,4,5$. Solutions 
are trivial in this uncoupled limit as the amplitude of the oscillators at 
each site can be chosen independently from the set 
$\psi_{n,m}\in\{0,\pm\sqrt{\Lambda/2Q_{3}}\}$, where we have made a 
restriction to real single-frequency solutions. From the configuration 
chosen in the anti-continuous limit it is possible to make a classification 
of the most fundamental localized solutions of Eq.~\eref{eq:xdnlsstat}. The 
set of solutions obtained from continuation to non-zero coupling of the 
configuration with one site excited and all other at rest will be called 
$1$-site DBs. Similarly, if two neighbouring sites are excited at the 
anti-continuous limit we will get symmetric or anti-symmetric $2$-site DBs, 
depending on the relative phases of the sites.

Since we ultimately are interested in the mobility of localized modes, we 
would like to compare the Hamiltonian of different stationary solutions. As 
has been described in Sec.~\ref{sec:introduction} (especially for the 
corresponding 1D model~\cite{Oster.03}), the enhanced mobility of localized 
modes is connected to the vanishing of a PN-barrier measuring the difference 
in Hamiltonian between fundamental stationary modes. The idea is that for a 
small or vanishing barrier, mobility of a highly localized mode may be 
thought of as a continuous transformation between, e.g., stationary $1$-site 
and $2$-site solutions, put into motion by adding energy to overcome the 
barrier (or pinning). Therefore, numerical continuation to non-zero coupling 
will be performed under the constraint of constant norm. As the parameters 
$Q_{2}$ and $Q_{3}$ are fixed from rescalings, the parameters $Q_{4}$ and 
$Q_{5}$ are varied, leaving the frequency $\Lambda$ as a free parameter to be 
determined by the stationary solution. The choice to keep $\mc{N}$ fixed is 
motivated by the fact that it is a dynamical parameter that is conserved 
under the evolution of Eq.~\eref{eq:xdnls}. Hence, when considering a range 
of coupling parameters (as will shortly be done), this will facilitate a 
meaningful comparison of the Hamiltonians over that entire range. This would 
not be the case had the continuation been performed for constant frequency. 
However, for a fixed frequency and varying norm the relevant energy quantity 
would instead be the grand canonical free-energy 
$\mc{G}=\mc{H}-\Lambda\mc{N}$, with $\Lambda$ acting as a chemical potential, 
as used in~\cite{Melvin.06}.

\subsection{Stability}
\label{sec:stability}
The linear stability of stationary modes is most simply investigated by 
studying the evolution of a small perturbation in the rotating frame of 
reference. Hence, for a given solution we make the ansatz 
$\Psi_{n,m}(z)=[\psi_{n,m}+(\epsilon_{n,m}+\rmi\eta_{n,m})\rme^{\lambda z}]
\rme^{-\rmi(\Lambda+Q_{1})z}$ in Eq.~\eref{eq:xdnls} and arrive at a linear 
matrix eigenvalue problem for the growth rates $\lambda$ of the real 
infinitesimal perturbations $\epsilon_{n,m}$ and $\eta_{n,m}$~\cite{Carr.85}. 
Since the problem is Hamiltonian this matrix is infinitesimally symplectic and 
will have the simultaneous eigenvalues $\pm\lambda$, 
$\pm\lambda^{*}$~\cite{Robinson.99}. Due to the global phase invariance, 
$\lambda=0$ is always a double eigenvalue. A stationary solution is linearly 
stable if all eigenvalues are located on the imaginary axis, i.e., we can 
actually at most make a statement about marginal stability of the solution 
(meaning that a perturbation at most will grow with a rate that is polynomial 
in $z$).

\begin{figure*}[t]
\begin{center}
   \includegraphics[width=\tw]{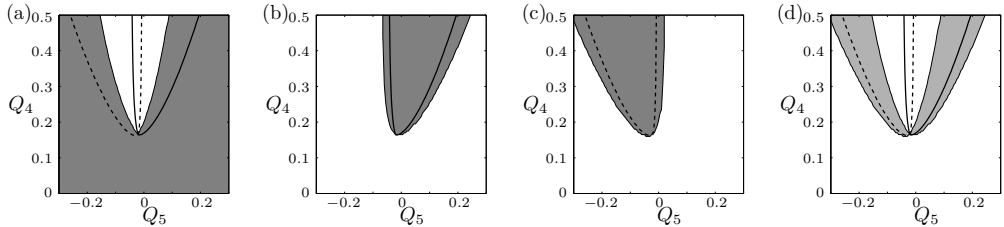}
\caption{Region of stability (shaded) for (a) $1$-site, 
(b) symmetric $2$-site and (c) anti-symmetric $2$-site solution for 
$Q_{2}=0.2$, $Q_{3}=0.5$ and $\mc{N}=5$; (d) region of existence (shaded) of 
intermediate solutions (unstable) between the stability boundaries of the 
$1$-site solution and the respective $2$-site solution 
[symmetric (anti-symmetric) in the right (left) region]. Additional small 
regions of stability, not shown here, can be found for the $2$-site solutions 
for relatively small $Q_{4}$. In the cases of multiple solutions, as for the 
$1$-site case, stability is indicated if at least one solution is stable. See 
also Figs.~\ref{fig:bifurc1} and~\ref{fig:bifurc2} for some details. The 
solid line in (a), (b) and (d) indicates where the energy difference, 
$\Delta\mc{H}=\mc{H}_{1}-\mc{H}_{2s}$, vanishes for the $1$-site and symmetric
$2$-site solutions. The dashed line in (a), (c) and (d) shows the same thing 
for the $1$-site and anti-symmetric $2$-site solutions, 
$\Delta\mc{H}=\mc{H}_{1}-\mc{H}_{2a}$. The general phenomenology is similar 
also for other values of the norm $\mc{N}$. The size of the lattice is 
$21\times 21$ with periodic boundary conditions.}
\label{fig:stabbound}
\end{center}
\end{figure*}
%
%
\begin{figure*}[t]
\begin{center}
   \includegraphics[width=\tw]{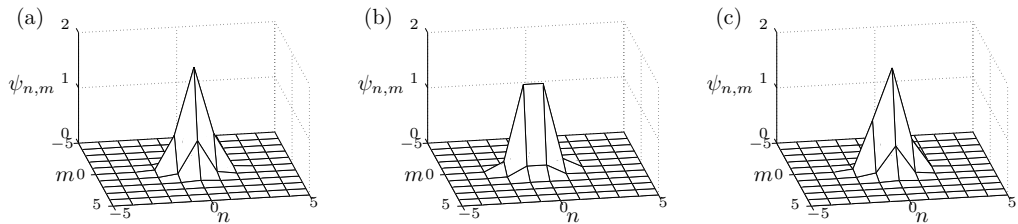}
\caption{Examples of solutions for 
$Q_{2}=0.2$, $Q_{3}=0.5$, $Q_{4}=0.25$, $Q_{5}=0.05$ and $\mc{N}=5$. 
(a) $1$-site solution with $\mc{H}=19.7435$ and $\Lambda=7.5019$; 
(b) symmetric $2$-site solution with $\mc{H}=19.9278$ and $\Lambda=7.6357$; 
(c) intermediate solution with $\mc{H}=19.7310$ and $\Lambda=7.5036$.}
\label{fig:sol}
\end{center}
\end{figure*}
\begin{figure*}[t]
\begin{center}
   \includegraphics[width=\tw]{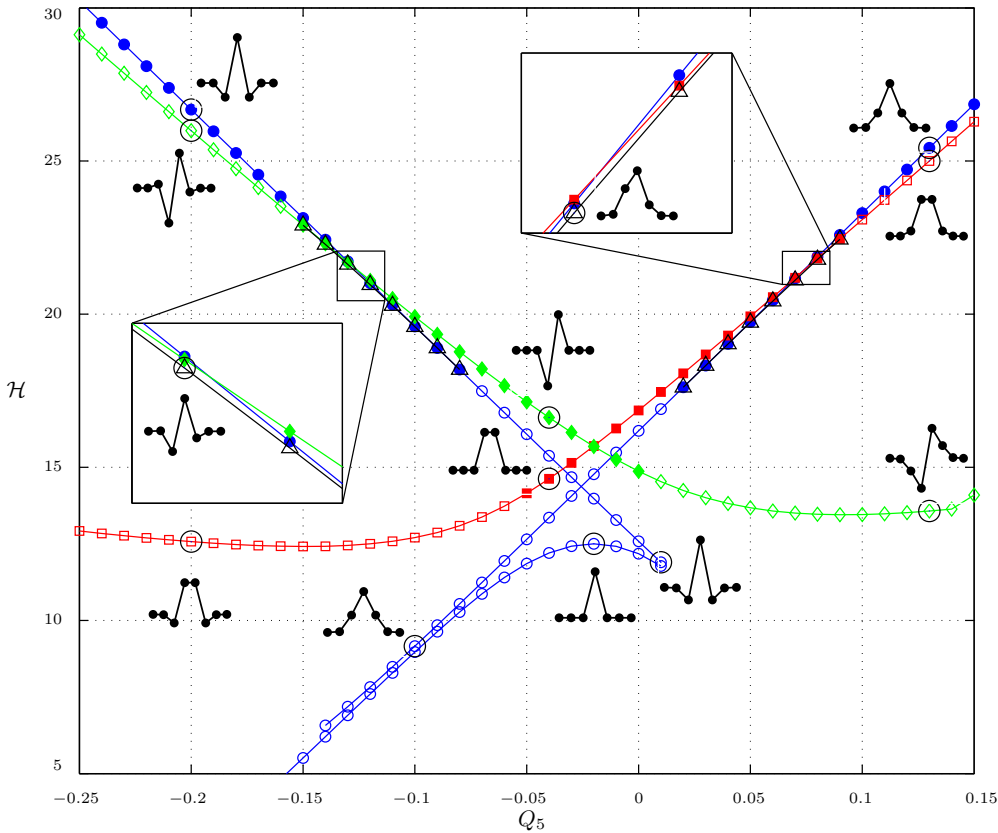}
\caption{(Color online) Bifurcation diagram with Hamiltonian $\mc{H}$ versus 
the parameter $Q_{5}$ for $Q_{2}=0.2$, $Q_{3}=0.5$, $Q_{4}=0.25$ and 
$\mc{N}=5$. $1$-site solutions are indicated with circles ($\circ$), symmetric 
$2$-site solutions with squares ($\square$), anti-symmetric $2$-site solutions 
with diamonds ($\Diamond$) and intermediate solutions with triangles 
($\triangle$). The markers are filled (unfilled) if the solution at that 
point is stable (unstable). Note also that a maximizer of the Hamiltonian 
always is stable. For some selected points (large circles) a one-dimensional 
cross section along a lattice direction of the two-dimensional solutions is 
shown to give a view of their characteristic form. The intermediate solutions 
are always unstable and exist only in a limited parameter range. They emerge 
from pitchfork bifurcations and connect the $1$-site solution with either of 
the $2$-site solutions. Note that they appear in connection with the vanishing 
of the energy difference (see insets).}
\label{fig:bifurc1}
\end{center}
\end{figure*}
\begin{figure}[t]
\begin{center}
   \includegraphics[height=0.4\tw, width=0.4\tw]{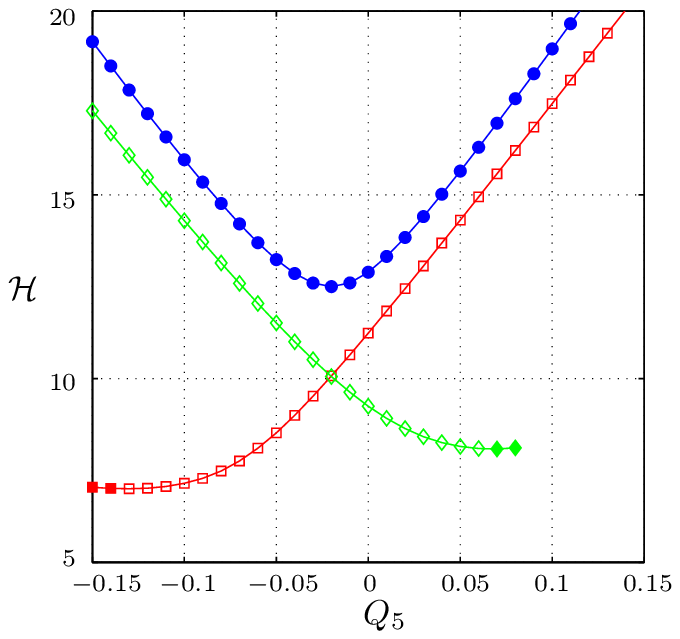}
\caption{(Color online) Same as in Fig.~\ref{fig:bifurc1} but for $Q_{4}=0.1$. 
Note here that $\mc{H}$ for the $1$-site solution is always larger than for 
the $2$-site solutions.}
\label{fig:bifurc2}
\end{center}
\end{figure}

Extensive calculations, covering a large portion of parameter space, have 
been done for the most fundamental stationary solutions. In 
Fig.~\ref{fig:stabbound} the regions of stability (shaded) for the $1$-site 
and the two $2$-site solutions are shown for fixed norm $\mc{N}=5$. For the 
$1$-site case there actually exist three different solutions in part of the 
region $Q_{4}>0.165$ (the edge of the instability region), each accessible by 
continuation but via different paths from the anti-continuous limit. Stability 
is indicated if one of the solutions is stable (there is no multistability of 
these three 1-site solutions and some details can also be found in 
Figs.~\ref{fig:bifurc1} and~\ref{fig:bifurc2}). The same behaviour appears 
also in 1D~\cite{Oster.03b}, and is presumably an effect of the nonlinear 
coupling terms that can lead to a vanishing effective coupling between lattice
sites. See further the discussion in Sec.~\ref{sec:compact}. From 
Fig.~\ref{fig:stabbound} we see that both stable and unstable solutions of all 
kinds exist and we note that the stability regions for the $2$-site solutions 
together cover the region of instability of the $1$-site solutions. For a 
quite large range of parameter values we thus have stable propagation of both 
types of solutions along the array of waveguides. The stability boundaries of 
the different solutions do not coincide. In the 
1D model~\cite{Oster.03,Oster.03b}, the stability boundaries very nearly 
coincide and in fact create a very narrow region of parameter space where the 
$1$-site solution and one of the $2$-site solutions have either simultaneous 
stability or instability. Moreover, the exchange of stability across this 
narrow range is connected to the existence of an intermediate asymmetric 
solution in this region, interpolating between the two solutions. When 
crossing 
the region of stability inversion, the intermediate solution appears from a 
pitchfork bifurcation as one of the solutions (say the $1$-site solution) 
changes its stability and then disappears in a reversed pitchfork bifurcation 
with the other ($2$-site) solution~\cite{Oster.03}. Here, in 2D, although the 
stability boundaries lie rather far apart in parameter space, unstable 
asymmetric intermediate solutions still exist between the stability boundaries 
of the $1$-site solutions and both $2$-site solutions 
(Fig.~\ref{fig:stabbound}d) and appear as the result of similar bifurcations. 
Examples of the solutions can be seen in Fig.~\ref{fig:sol}.

%
%
%
Moreover, comparing the Hamiltonians of the obtained solutions reveals the 
existence of zeros of the energy difference along some boundaries in parameter 
space. These have been indicated in Fig.~\ref{fig:stabbound}. To better 
clarify the relations among the different solutions, bifurcation diagrams for 
$Q_{4}=0.25$ and $Q_{4}=0.1$ are shown in 
Figs.~\ref{fig:bifurc1} and~\ref{fig:bifurc2}, with Hamiltonian $\mc{H}$ for 
the solutions plotted against the parameter $Q_{5}$. Note that the solution 
that maximizes $\mc{H}$ always is stable. From Fig.~\ref{fig:bifurc1} it is 
clear that the intermediate solution appears in connection with the vanishing 
energy difference, i.e., the difference in Hamiltonian, of the $1$-site and 
$2$-site solutions $\Delta\mc{H}=\mc{H}_{1}-\mc{H}_2$, and through 
bifurcations provide the mechanism of their exchange of stability. The 
intermediate solution always lies very close but has a slightly lower value of 
$\mc{H}$ and is therefore unstable, in contrast to the solutions in 1D that 
may be stable~\cite{Oster.03}. From a mobility point of view, it is the 
energy of the intermediate solution that must be overcome to achieve mobility. 
This additional energy (the PN-barrier) is, with reference to the stable 
modes, small compared to the total energy ($\lesssim 1\%$). Note also in 
Fig.~\ref{fig:bifurc1} that for each $2$-site solution there are two 
intersections with $1$-site solutions and hence two zeros of the energy 
difference. One is, as expected from the energy minimum principle, present in 
the region of existence of the intermediate solutions, where the $1$-site 
solution and the respective $2$-site solution have simultaneous stability and 
are the stationary modes with highest value of $\mc{H}$ 
(see the zoomed areas). The energy difference is also zero in a region where 
the involved solutions are not the ones with highest $\mc{H}$, and in 
particular the $1$-site solution is unstable in this region (see 
intersections at $\mc{H}\approx 15$). This intersection, as can be seen from 
the solution insets in Fig.~\ref{fig:bifurc1}, is with a solution possessing 
the `wrong' symmetry, i.e., the nearest neighbours of the centre site have 
the opposite sign compared to the symmetry or anti-symmetry of the respective 
$2$-site solution. Therefore exchange of stability, or mobility, cannot be 
expected at this vanishing of the energy difference.

In Fig.~\ref{fig:bifurc1} we also see that the branches of the $1$-site 
solutions terminate. Two of the branches will merge at $Q_{5}\approx 0.01$, 
thus the end of the branches are due to bifurcations. The behaviour of the 
three branches is not symmetric and the two branches lying close in the left 
part of the figure will not coincide. Instead, the upper branch of these two 
is terminated at $Q_{5}\approx -0.14$ in a bifurcation with a symmetric 
$5$-site solution that at the anti-continuous limit has $5$ in-phase 
neighbouring sites excited in a cross configuration. The lower branch can be 
continued to lower values of $Q_{5}$ where it will approach a plane wave 
solution.

\section{Mobility}
\label{sec:mobility}
A bit surprisingly, when considering the results of the stability analysis of 
stationary solutions, the mobility of localized modes in the 
model~\eref{eq:xdnls} is very poor and in stark contrast to the very excellent 
mobility in the 1D model~\cite{Oster.03} and the saturable model in 
1D~\cite{Hadzievski.04} and 2D~\cite{Vicencio.06} near points of minimal 
PN-barrier. The method used to try to induce mobility is to apply a phase 
gradient $\Psi_{n,m}\mapsto\Psi_{n,m}\rme^{\rmi(k_{x}n+k_{y}m)}$ across one 
of the stationary real solutions. In applications this can be achieved by 
launching the laser beam that excites the waveguides at an angle to the face 
of the array~\cite{Morandotti.01}. This perturbation will kick the solution 
in the direction of the phase 
gradient~\cite{Bang.96,Hadzievski.04,Vicencio.06,Oster.03,Maluckov.05} and 
will have an effect similar to the marginal mode perturbation at an 
instability boundary described in~\cite{Chen.96,Aubry.98}, which in a complex 
formulation will correspond to a pure imaginary (velocity) perturbation to the 
real (position) solution. The phase gradient perturbation also has the nice 
feature of preserving the norm. Here we are strictly interested in axial 
propagation as this is the direction of possible propagation in the lattice 
suggested by the minimal PN-barrier of the stationary modes. Therefore 
$k_{y}=0$ has consistently been used. Mobility in the diagonal direction 
would for the nearest-neighbour coupled lattice be connected to a vanishing 
energy difference for the $1$-site solution and a $4$-site solution 
(quadrupole). This does not appear in the present model and the $4$-site 
solutions generally has a much lower value of $\mc{H}$ than the $1$-site 
solution. The reason that it is possible in the model in~\cite{Vicencio.06} 
is the saturable nonlinearity that will limit the amplitude of the 
excitations, bringing the singly excited and multiply excited solutions 
closer in energy for high enough power. The same effect is also the reason for 
the multiple zeros of the energy difference, 
$\Delta\mc{H}=\mc{H}_{1}-\mc{H}_{2}$, in saturable media, as more and more 
excited sites reach the saturation limit. For the model~\eref{eq:xdnls} we 
should expect at most one zero of the PN-barrier with fixed coupling 
parameters and varying norm $\mc{N}$.

%
\begin{figure}[t]
\begin{center}
   \includegraphics[width=0.4\tw]{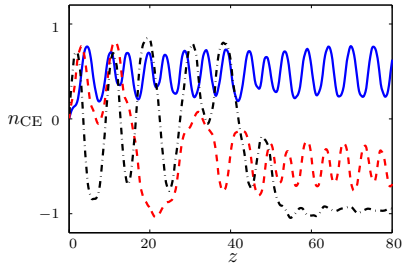}
\caption{(Color online) Evolution of the centre of energy when perturbing a 
$1$-site solution for $Q_{2}=0.2$, $Q_{3}=0.5$, $Q_{4}=0.25$, $Q_{5}=0.05$ 
and $\mc{N}=5$ (solution shown in Fig.~\ref{fig:sol}a) with a phase gradient 
$k_{x}$. In all cases $\mc{H}$ is sufficiently lowered to be below the value 
of the lowest stationary (the intermediate) solution with $\mc{H}=19.7310$. 
The solid line is for $k_{x}=0.1$ ($\mc{H}=19.7027$), the dotted line for 
$k_{x}=0.2$ ($\mc{H}=19.5817$) and the dash-dotted line for $k_{x}=0.5$ 
($\mc{H}=18.7839$).}
\label{fig:masscentre1}
\end{center}
\end{figure}
Trying to kick any of the stationary solutions in the region of existence of 
the intermediate solution, where the PN-barrier is minimal, will result in an 
oscillatory behaviour near the initial position. The centre of energy for the 
axial direction $\hat{\mathbf{n}}$, defined as
\begin{equation}
\label{eq:masscentre}
   n_{\rm{CE}} = \dfrac{1}{\mc{N}}\sum_{n,m} n\mc{N}_{n,m} \,,
\end{equation}
will at most be displaced about one lattice site. The generic behavour is 
examplified in Fig.~\ref{fig:masscentre1} where the $1$-site solution in 
Fig.~\ref{fig:sol}a is perturbed with different phase gradients. Although the 
perturbation is large enough to overcome the PN-barrier, set by the unstable 
intermediate solution, no mobility, except small displacements to neighbouring 
sites, is observed, indicating that the effect of pinning is stronger than 
suggested by the energy difference considerations. The final position of 
$n_{\rm{CE}}$ is not even always in the direction of the phase gradient kick, 
and is located either on or between sites corresponding to one of the stable 
stationary solutions. Since the initial velocity of the excitation given by 
the perturbation can be large, the absence of mobility here is not the effect 
of resonance with the multi-spectral bands appearing at low velocity for 
travelling waves near a zero of the PN-barrier~\cite{Melvin.06,Aigner.03}.

%
\begin{figure}[t]
\begin{center}
   \includegraphics[width=0.4\tw]{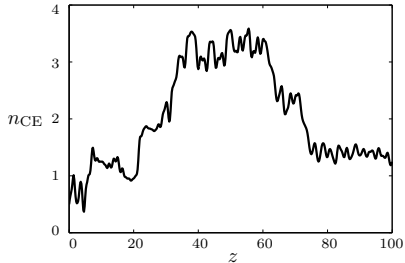}
\caption{Evolution of the centre of energy when perturbing an unstable 
symmetric $2$-site solution for 
$Q_{2}=0.2$, $Q_{3}=0.5$, $Q_{4}=0.25$, $Q_{5}=-0.06$ and $\mc{N}=5$ with a 
phase gradient $k_{x}=0.2$. The motion depends sensitively on initial 
conditions and boundary conditions. Here periodic boundary conditions on a 
$21\times 21$ lattice is used.}
\label{fig:masscentre2}
\end{center}
\end{figure}
If we instead start with solutions away from the region of existence of the 
intermediate solutions only small oscillations around the stable stationary 
solution in the present region is observed. Displacement of the centre of 
energy a few sites is however possible. In Fig.~\ref{fig:masscentre2} an 
unstable symmetric $2$-site solution is perturbed and the centre of energy 
moves up to two sites. But the irregular oscillations are due to a 
reconfiguration of the excitation from an initial more symmetric profile to, 
at the final position, a more anti-symmetric profile being closer to the 
stable stationary mode at the given values of the parameters. The motion is 
highly unpredictable and depends sensitively on the initial conditions and 
boundary conditions.

%
\begin{figure}[t]
\begin{center}
   \includegraphics[height=5cm, width=5cm]{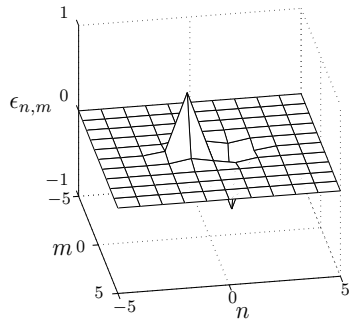}
\caption{Real part of the unstable eigenmode of the $1$-site solution for 
$Q_{2}=0.2$, $Q_{3}=0.5$, $Q_{4}=0.25$, $Q_{5}=0.0165$ and $\mc{N}=5$. By 
symmetry, the eigenmode is degenerate with a corresponding eigenmode oriented 
along the $\hat{\mathbf{m}}$ direction. The related eigenvalue is 
$\lambda=0.4140$.}
\label{fig:eigmode1}
\end{center}
\end{figure}
%
%
\begin{figure}[t]
\begin{center}
   \includegraphics[height=0.4\tw, width=0.4\tw]{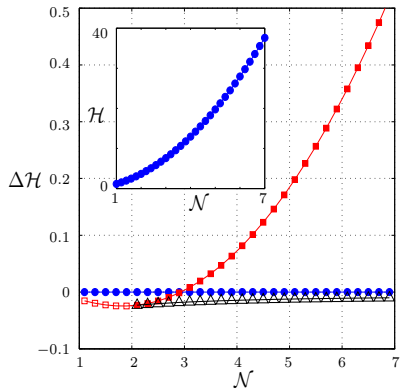}
\caption{(Color online) Hamiltonian $\mc{H}$, relative to the value of the 
$1$-site solution ($\circ$), versus norm $\mc{N}$ for $Q_{2}=0.2$, 
$Q_{3}=0.5$, $Q_{4}=0.25$ and $Q_{5}=0.05$. The intermediate solution 
($\triangle$) emerges from a pitchfork bifurcation with the symmetric 
$2$-site solution ($\square$) at $\mc{N}\approx 2.1$. It is always the 
solution with the slightly lower value of $\mc{H}$ and is unstable. As 
$\mc{N}$ increases, although very close in $\mc{H}$, there will be no 
bifurcation with the $1$-site solution, implying that the intermediate 
solution will not connect the $1$-site and the $2$-site solutions when 
considered as a function of the dynamical parameter $\mc{N}$. The inset shows 
$\mc{H}$ versus $\mc{N}$ for the $1$-site solution, as a reference.}
\label{fig:bifurc3}
\end{center}
\end{figure}
%
%
\begin{figure}[t]
\begin{center}
   \includegraphics[height=0.4\tw, width=0.4\tw]{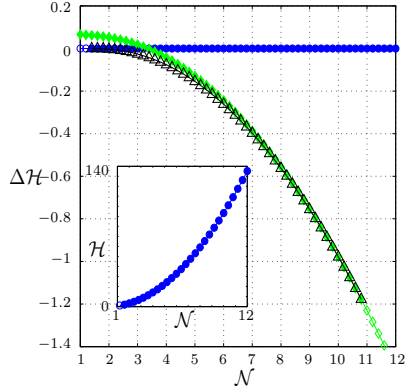}
\caption{(Color online) Hamiltonian $\mc{H}$, relative to the value of the 
$1$-site solution ($\circ$), versus norm $\mc{N}$ for $Q_{2}=0.2$, 
$Q_{3}=0.5$, $Q_{4}=0.25$ and $Q_{5}=-0.14$. The intermediate solution 
($\triangle$) appears at $\mc{N}\approx 1.4$ and disappears at 
$\mc{N}\approx 10.9$ in a bifurcation with the anti-symmetric $2$-site 
solution ($\Diamond$). The inset shows $\mc{H}$ versus $\mc{N}$ for the 
$1$-site solution, as a reference.}
\label{fig:bifurc5}
\end{center}
\end{figure}
%
%
\begin{figure}[t]
\begin{center}
   \includegraphics[height=0.4\tw, width=0.4\tw]{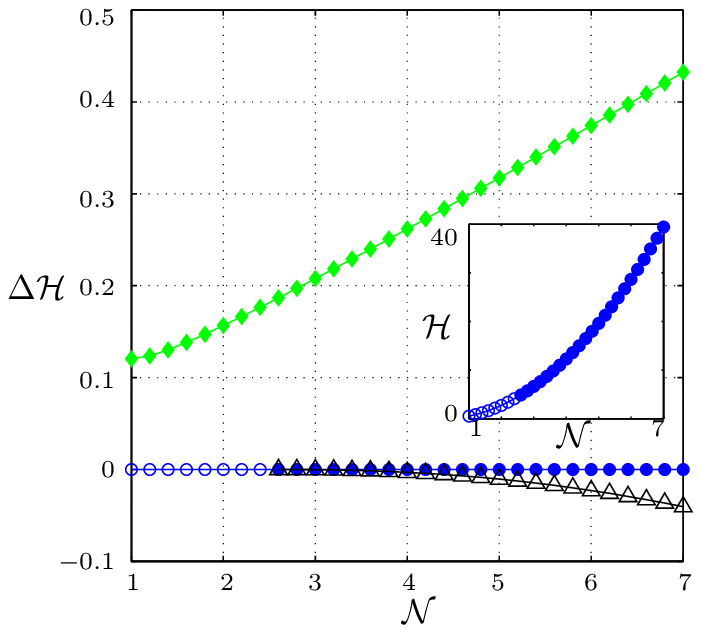}
\caption{(Color online) Hamiltonian $\mc{H}$, relative to the value of the 
$1$-site solution ($\circ$), versus norm $\mc{N}$ for $Q_{2}=0.2$, 
$Q_{3}=0.5$, $Q_{4}=0.25$ and $Q_{5}=-0.1$. In this case the intermediate 
solution ($\triangle$) emerges from a pitchfork bifurcation with the $1$-site 
solution at $\mc{N}\approx 2.4$ and there is actually no zero of the energy 
difference. The anti-symmetric $2$-site solution ($\Diamond$) always has a 
higher value of $\mc{H}$ and is stable. The inset shows $\mc{H}$ versus 
$\mc{N}$ for the $1$-site solution, as a reference.}
\label{fig:bifurc4}
\end{center}
\end{figure}
At the edge of the stability boundary, where the bifurcation with the 
intermediate solution occurs for the fundamental stationary modes, the 
eigenmodes related to the real eigenvalues that are responsible for the 
instability has a form that may be indicative of mobility. The real part of 
this growing mode is shown in Fig.~\ref{fig:eigmode1} for the $1$-site 
solution. Exactly at the stability boundary the imaginary part of the 
eigenmode is zero. From the form of the eigenmode we conclude that perturbing 
along this mode (`depinning mode' in~\cite{Chen.96}) will promote a change of 
shape of the $1$-site solution towards a profile that resembles the symmetric 
$2$-site solution (or the intermediate solution). The opposite relation holds 
at the stability boundary of the $2$-site solution. Hence the form of the 
growing modes supports the heuristic view of mobility of narrow excitations as 
a transformation between stationary solutions 
(see also Fig.~3 in~\cite{Oster.03}). But in the present model the stability 
boundaries are far apart in parameter space and such a transformation may not 
be possible. To be conclusive in this matter we will need to fix the coupling 
parameters and investigate the exchange of stability as a function of the 
dynamical parameters $\mc{H}$ and $\mc{N}$. If $\mc{N}$ is varied for a set 
of fixed coupling parameters inside the right region of existence of the 
intermediate solution in Fig.~\ref{fig:bifurc1} we can find a situation when 
the intermediate solution will only bifurcate with the symmetric $2$-site 
solution, and not with the $1$-site solution for reasonable values of the norm 
($\mc{N}<100$). This is illustrated in Fig.~\ref{fig:bifurc3}. In the event 
that a bifurcation occurs with both the fundamental solutions involved 
(Fig.~\ref{fig:bifurc5}), the bifurcation points are far apart measured also 
in terms of the dynamical parameters. Measuring the distance between the 
bifurcation points in norm ($\Delta\mc{N}$) we find that it is always 
comparable in size to the value of the norm at the bifurcation point with the 
higher value, i.e., $\Delta\mc{N}/\mc{N}\sim 1$. The PN-barrier 
($\Delta\mc{H}$) is small compared to $\mc{H}$ for all values of the norm. 
Since the bifurcation points are either far apart or the intermediate solution 
does not connect the $1$-site and $2$-site solutions at all, this indicates 
that there is no way to transform between the two. Put in another way, this 
suggests that there is no trajectory in phase space that passes close to the 
stationary configurations and consequently the mobility is very poor. Our 
numerical observations lead us to the conclusion that mobility is resisted 
since the site-centred and bond-centred configurations are too far apart for 
the adjustment of $\mc{N}$ and $\mc{H}$ to be possible by exchange with an 
accompanying oscillating background, or in this case mainly an adjustment of 
$\mc{N}$ since the difference in energy is always relatively small.  A 
tractable analysis could possibly be made along the lines of the method of an 
effective Hamiltonian where the dynamics is averaged over periodic 
trajectories in phase space~\cite{MacKay.02}. The more explicit interpretation 
of the PN-barrier as referred to a trajectory in phase space would definitely 
be more suitable in the present context. However, some extensions are 
necessary as an evaluation of the method shows that it gives an accurate 
description of mobility only close to limits where the travelling excitations 
are exact~\cite{MacKay.02}. But this does not exclude that some insight might 
be given to the effect of pinning observed here. Further, as shown in 
Fig.~\ref{fig:bifurc4}, there are cases when the PN-barrier does not vanish at 
all, but the intermediate solution still exists.

\section{Compact solutions}
\label{sec:compact}
An interesting effect of the nonlinear coupling terms in Eq.~\eref{eq:xdnls} 
is that they may cancel the linear coupling and effectively give rise to a 
total coupling that is zero~\cite{Oster.03,Oster.05}. This property is also 
the criterion for the existence of exact compact solutions having strictly 
zero amplitude outside an interval~\cite{Kevrekidis.02,Kevrekidis.03}. The 
criterion will take the form of a parameter constraint and this implies that 
the compact solutions, as exact mathematical entities, are not robust with 
respect to parameter variation. They are interesting anyway from a 
physical perspective, since a near perfect localization prevails in the 
neigbourhood of the exact parameter values, and this family of 
strongly localized solutions may 
very well be dynamically stable. The solution will develop an exponentially 
decaying tail, but the contrast, i.e., the ratio of the amplitude of 
neighbouring sites to the amplitude of the excited sites, is still large which 
can be desirable for applications~\cite{Bang.96,Oster.03}. Further, although 
we speak of exact compact solutions in the context of the lattice 
equation~\eref{eq:xdnls}, this does not imply compactness in the real system. 
The equation describes the amplitude of the modes in the respective 
waveguides, but the modes are themselves spatially exponentially decaying. The 
extreme localization represented by compact solutions will nonetheless be 
preferable if several excitations propagate in the system at the same time as 
these will constitute excitations with least mutual interaction. The existence 
of compact solutions has an impact on the properties of the system under study.

The idea of the compactification is that for a particular amplitude the total 
coupling can vanish when the neighbouring sites have zero amplitude, leading 
to a decoupling of the lattice between these sites. Take, e.g., the $1$-site 
excitation with only the site $(n,m)$ excited and all others at rest and put 
this ansatz into Eq.~\eref{eq:xdnlsstat}. From the equation on one of the 
neighbouring sites, e.g., site $(n-1,m)$, we get 
$Q_{2}+2Q_{5}|\psi_{n,m}|^{2} = 0$. For a compact solution with $M$ excited 
sites, this constraint generalises to
\begin{equation}
\label{eq:Qeff}
   Q_{2}+2Q_{5}\dfrac{\mc{N}}{M} = 0,
\end{equation}
if all sites have the same amplitude. Note that $Q_{5}<0$ is required for the 
existence of compact solutions (as $Q_{2}>0$). This criterion corresponds to 
having a negative Kerr index of the nonlinear material surrounding the 
waveguides, or for BEC in an optical lattice an effective interatomic 
attraction in the dilute gas (as for $^{7}$Li). Solving for the $1$-site and 
$2$-site compact solutions the results are the same as in 
1D~\cite{Oster.03,Oster.05}. From Eq.~\eref{eq:Qeff} the constraint for the 
$1$-site solution will correspond to the line $Q_{5}=-0.02$ in 
Fig.~\ref{fig:stabbound}, that precisely cuts through the edge of the 
instability region of the $1$-site solution and hence also seperates the two 
different regions of existence of the intermediate solutions. The loose 
interpretation of this constraint as an effective coupling is further 
strenghtened when considering that the two $2$-site solutions will have the 
same value of $\mc{H}$ along the plane of existence of the $1$-site solution. 
Note, e.g., that the lines of $\mc{H}$ intersect at $Q_{5}=-0.02$ in the 
bifurcation diagrams in Figs.~\ref{fig:bifurc1} and~\ref{fig:bifurc2}.

Compact solutions with more than two sites excited can also be found. There 
are two different classes of real $3$-site solutions, with the center site 
being either in-phase or out-of-phase with the two outermost sites. Having, 
in 2D, the freedom to choose the orientation of the three sites in the lattice 
as being all along one lattice direction or in an `L'-configuration, the 
presence of coupling in two different lattice directions will also mean less 
freedom when compared to the 1D system. The extra restrictions imposed for the 
$3$-site solutions implies that all sites must have the same amplitude, which 
is not required in 1D. Explicit calculations yield the same frequency 
$\Lambda=2Q_{3}\mc{N}/3$ for all solutions, and apart from the parameter 
constraint~\eref{eq:Qeff} also the constraint $Q_{4}=\mp Q_{5}$, where the 
upper sign is for the in-phase solution and the lower sign for the 
out-of-phase solution. A check of the stability shows that only the 
out-of-phase solution will be stable for $\mc{N}\geq 8.72$, regardless of the 
orientation in the lattice. It is also possible to solve analytically for 
compact $4$-site and $5$-site solutions, some of which are stable for large 
enough values of the norm ($\mc{N}\gtrsim 20$).

\subsection{Vortices and complex solutions}
\label{sec:complex}
\begin{figure}[t]
\begin{center}
   \includegraphics[width=0.45\tw]{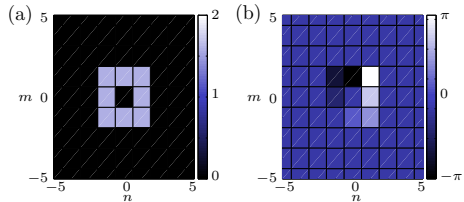}
\caption{(Color online) (a) The amplitude $\sqrt{\mc{N}_{n,m}}=|\Psi_{n,m}|$ 
and (b) the phase $\theta_{n,m}=\textrm{arg}\{\Psi_{n,m}\}$ for a stable 
stationary compact vortex of topological charge $S=1$ for 
$\mc{N}=20$, $Q_{2}=0.2$, $Q_{3}=0.5$, $Q_{4}=0.04$ and 
$Q_{5}=-4Q_{2}/\mc{N}=-0.04$. The frequency is given by 
$\Lambda=-3\sqrt{2}Q_{2}+Q_{3}\mc{N}/4+Q_{4}\mc{N}=2.4515$.}
\label{fig:vortex}
\end{center}
\end{figure}

An interesting extension when going from one to two spatial dimensions is the 
possiblity to support localized excitations having a topological charge, which 
is proportional to the angular momentum carried by the excitation. We define 
the topological charge $S$ as the number of complete $2\pi$ twists of the 
phase when moving on a contour around the excitation. Such discrete vortex 
solitons have been observed in experiments~\cite{Neshev.04,Fleischer.04} and 
described in the context of DNLS-type 
equations~\cite{Johansson.98,Malomed.01,Pelinovsky.05,Oster.06}. In the 
model~\eref{eq:xdnls} \emph{compact vortices} can be supported. The 
constraint~\eref{eq:Qeff} must still be fulfilled, which means, as we have 
fixed $Q_{2}$ and $Q_{3}$ by rescalings, that the parameter $Q_{5}$ is fixed 
if the norm $\mc{N}$ and $Q_{4}$ are used as free parameters. The most simple 
compact vortex is given by four excited sites in a square configuration with a 
phase difference of $\pi/2$ between neighbouring sites giving a topological 
charge $S=1$. This solution will have the frequency 
$\Lambda=Q_{3}\mc{N}/2+Q_{4}\mc{N}$ and have a very small region of stability 
near $Q_{4}=0$ (e.g., for $\mc{N}=10$ it is stable for $Q_{4}\leq 0.0023$). 
For the square $4$-site configuration placed diagonally in the lattice, with 
a zero-amplitude mediating site in the middle, the frequency is instead 
$\Lambda=Q_{3}\mc{N}/2$ and the region of stability is dramatically extended 
so that for $\mc{N}=10$ the configuration is stable for $Q_{4}\leq 0.1634$. 
The excited sites are not nearest neighbours in this case and will only weakly 
interact. In fact for $\mc{N}=20$ the instability will set in at 
$Q_{4}=0.165$, which exactly corresponds to the instability of the compact 
$1$-site solution for $\mc{N}=5$. With more excited sites it is possible to 
find larger $S=1$ vortices that are stable for certain parameter values. In 
Fig.~\ref{fig:vortex} an example with eight sites is shown. Also higher order 
compact vortices exist. The configuration with eight sites on a square contour 
with phase difference $\pi/2$ between the sites will have charge $S=2$ and  
frequency $\Lambda=Q_{3}\mc{N}/4+Q_{4}\mc{N}/2$. It will also be stable for 
large norm ($\mc{N}\gtrsim 20$) and small $Q_{4}$. (Note that, by contrast, 
this $S=2$ configuration is always \emph{unstable} in the standard 
on-site DNLS model~\cite{Pelinovsky.05}.)
 
%

The vortex solutions are necessarily complex solutions as they have a phase 
difference between neighbouring sites other than $0$ or $\pi$, i.e., they 
cannot be transformed to real form using the global phase invariance of the 
DNLS-type equations. This implies that the norm current, Eq.~\eref{eq:JN}, is 
non-zero and that there is a flow of norm around the contour of the vortex 
balanced according to the discrete continuity equation~\eref{eq:Neq} with 
constant norm density $\mc{N}_{n,m}$. As described for the 1D model 
in~\cite{Oster.05} the presence of the nonlinear coupling will introduce a 
non-trivial phase dependence in the norm current density. In particular, the 
norm current $\mc{J}_{n,m}^{(n)}$ in Eq.~\eref{eq:JN} can become zero for 
specific values of the phase difference given by
\begin{equation}
\label{eq:cosphi}
   \cos\varphi_{n,m} =
      -\dfrac{Q_{2}+2Q_{5}(\mc{N}_{n,m}+\mc{N}_{n+1,m})}{
         4Q_{4}\sqrt{\mc{N}_{n,m}\mc{N}_{n+1,m}}}.
\end{equation}
Thus under the constraint~\eref{eq:cosphi} stationary complex solutions can 
exist without having to form a closed loop for the norm current flow, which 
is the case for a vortex, simply because there is no current. The compact 
complex $2$-site solution present in the 1D model~\cite{Oster.05} exists also 
here, with $\cos\varphi_{n,m}=Q_{2}/2Q_{4}\mc{N}$ and 
$\Lambda=Q_{3}\mc{N}+Q_{4}\mc{N}-3Q_{2}^{2}/2Q_{4}\mc{N}$. It is also stable 
for a range of phase twists, e.g., for $\mc{N}=10$ stability is found for 
$0.010\leq Q_{4}\leq 0.0420$, equivalent to 
$0.2381\leq\cos\varphi_{n,m}\leq 1$. Also compact complex solutions with more 
sites excited can be constructed, e.g., $4$-site non-vortex solutions that 
generally seem to be unstable. Neither these solutions nor the vortex 
solutions above need to be compact, but can be continued to non-compact 
solutions by lifting the constraint set by Eq.~\eref{eq:Qeff}. Although this 
may be an interesting further development the given examples serve to give an 
idea of the phenomena at hand.

\section{Power transport by plane waves}
\label{sec:planewave}
Leaving the localized solutions, we will in this section focus on the 
transport properties of plane waves. The excitation of the lattice with a 
plane wave correponds to an even distribution of the energy over all sites in 
the lattice.  A travelling wave on the form
\begin{equation}
\label{eq:pw}
   \Psi_{n,m}(z) = \sqrt{\rho}\,\rme^{-\rmi(\varphi n+\phi m +\Lambda z)}
\end{equation}
will have a norm density (power per site) $\mc{N}_{n,m}\equiv\rho$ and be 
a solution of Eq.~\eref{eq:xdnls} for the frequency
\begin{equation}
\begin{split}
\label{eq:pwfreq}
   \Lambda = & \phantom{0} 2Q_{3}\rho +
2(Q_{2}+8Q_{5}\rho)(\cos\varphi+\cos\phi) \\
      & +4Q_{4}\rho(4+\cos 2\varphi+\cos 2\phi).
\end{split}
\end{equation}
Note that the dispersion relation~\eref{eq:pwfreq} can be derived from 
Eq.~\eref{eq:Heq_theta} in the form 
$\Lambda=\partial{\mc{H}_{n,m}}/\partial\rho$. Though being a stationary 
solution, the plane wave~\eref{eq:pw} is associated with a balanced transfer 
of norm (power) between the sites due to the induced phase difference 
(wave number). This balance is described by the discrete continuity 
equation~\eref{eq:Neq} with constant norm density. From the interpretation of 
$\mc{J}^{(n)}\equiv\mc{J}_{n,m}^{(n)}$ and 
$\mc{J}^{(m)}\equiv\mc{J}_{n,m}^{(m)}$ as the norm transferred from a given 
site to its neighbours in the positive $\hat{\mathbf{n}}$ and 
$\hat{\mathbf{m}}$ directions we can, from Eq.~\eref{eq:JN}, obtain the norm 
current flowing in the lattice for a plane wave solution,
\begin{equation}
\begin{split}
\label{eq:Jvec}
   \boldsymbol{\mc{J}} = 
   & \mc{J}^{(n)}\hat{\mathbf{n}}+\mc{J}^{(m)}\hat{\mathbf{m}}
   = -\dfrac{\partial\mc{H}_{n,m}}{\partial\varphi}\hat{\mathbf{n}}
      -\dfrac{\partial\mc{H}_{n,m}}{\partial\phi}\hat{\mathbf{m}} \\
   = & \phantom{0} 2\rho\sin\varphi
      \big[Q_{2}+4Q_{4}\rho\cos\varphi+4Q_{5}\rho\big]\hat{\mathbf{n}} \\
   & + 2\rho\sin\phi
      \big[Q_{2}+4Q_{4}\rho\cos\phi+4Q_{5}\rho\big]\hat{\mathbf{m}}.
\end{split}
\end{equation}
Further, it is convenient to introduce a polar representation of the current 
defined by the relation 
$|\boldsymbol{\mc{J}}|\rme^{\rmi\alpha}=\mc{J}^{(n)}+\rmi\mc{J}^{(m)}$, 
so that
\begin{equation}
\label{eq:Jmag}
   |\boldsymbol{\mc{J}}| = \sqrt{\mc{J}^{(n)\,2}+\mc{J}^{(m)\,2}}
\end{equation}
is the magnitude of the current and
\begin{equation}
\label{eq:Jdir}
   \alpha = \arctan\left(\dfrac{\mc{J}^{(m)}}{\mc{J}^{(n)}}\right)
      \, (\pm\pi) \quad \in \, ]-\pi,\pi[
\end{equation}
its direction measured counter clockwise relative to the 
$\hat{\mathbf{n}}$ direction. Note that the power transported along the 
waveguides is proportional to $\rho$ and that the 
quantities~\eref{eq:Jvec}, \eref{eq:Jmag} and~\eref{eq:Jdir} describe the 
transport across the array. For a finite array such transport can only be 
sustained if there is a supply of power at one boundary and a drain at the 
opposite boundary.

\subsection{Modulational stability}
\label{sec:MI}
To ensure that the plane wave~\eref{eq:pw} has stable propagation its 
behaviour under a modulation is investigated. Adding a modulating 
perturbation with wave numbers $q$ and $p$ we make the ansatz
\begin{equation}
\begin{split}
\label{eq:MIansatz}
   \Psi_{n,m}(z) = &
      \big[\sqrt{\rho}
      +u(z)\rme^{\rmi(qn+pm)}+v^{*}(z)\rme^{-\rmi(qn+pm)}\big] \\
      & \times\exp[-\rmi(\varphi n+\phi m +\Lambda z)]
\end{split}
\end{equation}
in 
Eq.~\eref{eq:xdnls}~\cite{Oster.05,Smerzi.03,Smerzi.03b,Menotti.03,Kivshar.92}.
 Keeping only terms linear in $u$ and $v$, the modulational stability can be 
determined from the equations
\begin{equation}
   \rmi\dfrac{d}{dz}\left(\atop{u}{v}\right)
      = \left(\atop{a+b}{-c}\atop{c}{a-b}\right)
\left(\atop{u}{v}\right)
      = \omega_{\pm}\left(\atop{u}{v}\right) \,,
\end{equation}
with
\begin{subequations}
\begin{align}
   a = & \phantom{0}
      2(Q_{2}+8Q_{4}\rho\cos\varphi+8Q_{5}\rho)
         \sin\varphi\sin q \nonumber \\
      & +2(Q_{2}+8Q_{4}\rho\cos\phi+8Q_{5}\rho)
         \sin\phi\sin p \,, \\[2mm]
   b = & \phantom{0}
      2Q_{3}\rho - 2Q_{2}(\cos\varphi+\cos\phi) \nonumber \\
      & +4Q_{4}\rho[2\cos q(1+\cos 2\varphi)-\cos 2\varphi] \nonumber \\
      & +4Q_{4}\rho[2\cos p(1+\cos 2\phi)-\cos 2\phi] \nonumber \\
      & +2(Q_{2}+8Q_{5}\rho)(\cos\varphi\cos q + \cos\phi\cos p) \,, \\[2mm]
   c = & \phantom{0}
      2Q_{3}\rho
      +4Q_{4}\rho(\cos 2\varphi+2\cos q + \cos 2\phi+2\cos p) \nonumber \\
      & +8Q_{5}\rho[\cos\varphi(1+\cos q) + \cos\phi(1+\cos p)] \,.
\end{align}
\end{subequations}
The plane wave will be stable if the eigenfrequencies 
$\omega_{\pm}=a\pm\sqrt{b^{2}-c^{2}}$ are real for all $q,p\in]-\pi,\pi]$. 
By explicit calculation we find that the eigenfrequencies can be written on 
the form
\begin{align}
\label{eq:weig}
   \omega_{\pm} = & \phantom{0}
      \dfrac{\partial\mc{J}^{(n)}}{\partial\rho}\sin q
      + \dfrac{\partial\mc{J}^{(m)}}{\partial\rho}\sin p \nonumber \\
      & \pm\Bigg[
      \Bigg(\dfrac{4\sin^{2}\frac{q}{2}}{m_{\mc{H}}^{(n)}(\varphi)}
      +\dfrac{4\sin^{2}\frac{p}{2}}{m_{\mc{H}}^{(m)}(\phi)} \Bigg) \nonumber\\
      & \phantom{\pm\Bigg[}
      \times\Bigg(f(\varphi)\sin^{2}\frac{q}{2}+f(\phi)\sin^{2}\frac{p}{2}
      +\rho\dfrac{\partial\Lambda}{\partial\rho} \Bigg)\Bigg]^{1/2}\!\!,
\end{align}
where we have introduced the (energetic) effective mass
\begin{align}
\label{eq:mass}
   \dfrac{1}{m_{\mc{H}}^{(n)}}
   & = \dfrac{1}{\rho}\frac{\partial^{2}\mc{H}_{n,m}}{\partial\varphi^{2}}
   = -\dfrac{1}{\rho}\frac{\partial\mc{J}^{(n)}}{\partial\varphi} \nonumber \\
   & = -2(Q_{2}+4Q_{5}\rho)\cos\varphi-8Q_{4}\rho\cos 2\varphi \,,
\end{align}
with a similar expression for $m_{\mc{H}}^{(m)}(\phi)$, and the function
\begin{equation}
\label{eq:f}
   f(\varphi) = -2[(Q_{2}+12Q_{5}\rho)\cos\varphi 
      + 4Q_{4}\rho(2\cos^{2}\varphi+1)],
\end{equation}
entirely in correspondence with the result in 1D~\cite{Oster.05}. From the 
first factor in Eq.~\eref{eq:weig} we see that if the effective masses have 
opposite signs there will be at least some values of $q$ and  $p$ that will 
yield instability. An important observation that follows is that if 
$1/m_{\mc{H}}^{(n)}$ or $1/m_{\mc{H}}^{(m)}$ change sign due to tuning of some 
available parameters, the plane wave can only change its stability or remain 
unstable (unless the second factor in Eq.~\eref{eq:weig} changes sign at the 
same time).

\subsection{Power currents}
\label{sec:power}
Generally, the direction of power flow in a lattice will be completely 
determined by the induced phase gradient of the plane wave. For the cubic 
on-site DNLS equation ($Q_{4}=Q_{5}=0$), e.g., the direction of the norm 
current is fixed by the relative phases between the sites, 
$\tan\alpha\propto\sin\phi/\sin\varphi$, while the magnitude is linear in the 
amplitude, $|\boldsymbol{\mc{J}}|\propto\rho$. Thus any control of the norm 
current by tunable parameters is restricted to the wave numbers of the plane 
wave. With current experimental techniques, a phase gradient is induced by 
having the exciting laser beam directed at an angle to the array of 
waveguides~\cite{Morandotti.01}, so the wave numbers are easily adjustable but 
not rapidly tuned. Also, for a finite array the wave numbers will be quantized 
in order to satisfy the imposed boundary conditions.

As discussed in~\cite{Oster.05}, with the introduction of nonlinear coupling 
terms in the model~\eref{eq:xdnls}, the norm current will have a nonlinear 
dependence on the amplitude. This will make $\rho$ more suitable as a tunable 
parameter, also from an application point of view. Including the $Q_{5}$ term, 
but keeping $Q_{4}=0$ (which would be relevant for application to BEC in an 
optical potential~\cite{Smerzi.03,Smerzi.03b,Menotti.03}), will introduce a 
nonlinear dependence in the magnitude $|\boldsymbol{\mc{J}}|$ through the 
factor $Q_{2}+4Q_{5}\rho$. The sign of this factor will determine in which of 
the two possible directions ($\alpha$), differing by an angle $\pi$ and fixed 
by the phase gradients $\varphi$ and $\phi$, the current will flow. For 
$Q_{5}<0$ it is hence possible to make the current zero for a finite value of 
the amplitude $\rho$ and at the same time reverse its direction, just as in 
1D~\cite{Oster.05}. However, as the factor $Q_{2}+4Q_{5}\rho$ appears also in 
the effective mass~\eref{eq:mass}, the change of sign will affect the 
stability of the travelling plane wave. In particular, as noted in 
Sec.~\ref{sec:MI}, the plane wave cannot be modulationally stable for both 
directions of the current near the point of inversion. 

%
\begin{figure}[t]
\begin{center}
   \includegraphics[width=0.4\tw]{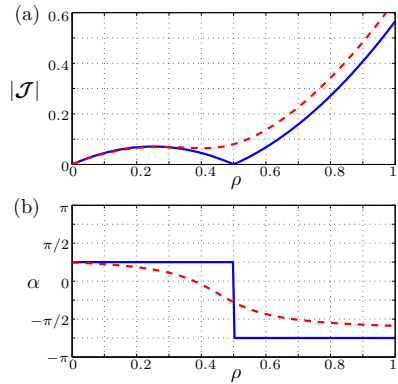}
\caption{(Color online) (a) The magnitude $|\boldsymbol{\mc{J}}|$ and (b) the 
direction $\alpha$ of the norm current for varying amplitude $\rho$ with the 
parameters $Q_{2}=0.2$, $Q_{3}=0.5$, $Q_{4}=0.1$ and $Q_{5}=-0.1$. The solid 
line is for $\varphi=\phi=\pi/2$, and the plane wave is stable for all $\rho$. 
The dashed line is for $\varphi=1.3$ and $\phi=1.9$, and the plane wave is 
stable for $\rho>0.119$.}
\label{fig:current}
\end{center}
\end{figure}
Inclusion also of the $Q_{4}$ term will change this situation, as the zeros of 
$\mc{J}^{(n)}$ and 
$1/m_{\mc{H}}^{(n)}\propto\partial\mc{J}^{(n)}/\partial\varphi$ in general no 
longer will coincide. However, numerical investigations reveal that stable 
inversion is not possible in the axial direction, i.e., with $\phi=0, \pi$, 
for any values of $\varphi$ and $Q_{4}>0$, but it is possible in other 
directions. Taking $\varphi=\phi\neq 0, \pi$ will give a diagonal direction of 
the current, with $\alpha=\pi/4,-3\pi/4$ depending on the relative signs of 
$\mc{J}^{(n)}$ and $\mc{J}^{(m)}$, and always stability near the point of 
inversion. An example of such a flip of direction as the amplitude of the plane 
wave is varied is shown in Fig.~\ref{fig:current}. An even greater 
flexibility is presented if different phase gradients are chosen for the 
$\hat{\mathbf{n}}$ and $\hat{\mathbf{m}}$ directions, $\varphi\neq\phi$. As 
illustrated in Fig.~\ref{fig:current} we may have the remarkable situation 
that as the amplitude $\rho$ is tuned the magnitude of the current is near 
constant while \emph{the direction is continuously changed} an angle $\pi/2$ 
over a stable range. Such an exact control of the power flow by simply 
changing the amplitude of the plane wave may find useful applications. One 
possible application could be to use the array of waveguides, excited with a 
plane wave, as the link in a power coupling device. The advantage over a 
direct coupling would be a greater control of the flow facilitated by the 
control of the excitation in the waveguides. The effect of current inversion 
could form the basis for an amplitude-operated power-switch. However, any 
discussions in this direction other than mere speculative are beyond the scope 
of this paper.

\section{Conclusion}
\label{sec:conclusion}
In conclusion, we have studied the extension to two spatial dimensions of a 
one-dimensional model describing coupled optical waveguides going beyond the 
lowest-order approximation and including nonlinear coupling, which, e.g., is 
relevant for arrays of linear waveguides embedded in nonlinear media. 
The phenomenology of the 1D model presented in a series of 
papers~\cite{Oster.03,Oster.05,Oster.03b} has been revisited in the 2D 
setting. Especially, we have shown that a vanishing energy difference between 
$1$-site and $2$-site stationary solutions and a subsequent small 
Peierls-Nabarro energy barrier, taking into account also the existence of 
asymmetric intermediate solutions, does \emph{not} result in good mobility of 
highly localized modes. The main reason, and the difference to models that 
show good mobility~\cite{Hadzievski.04,Vicencio.06,Oster.03}, is that, 
despite small energy differences, the stationary solutions are still in some 
sense far apart. Rather, the bifurcation points where the fundamental modes 
change their stability and where they exhibit a depinning mode promoting 
mobility are, although close in Hamiltonian ($\mc{H}$), far apart in norm 
($\mc{N}$) (Figs.~\ref{fig:bifurc3} and~\ref{fig:bifurc5}). From this we 
conclude that for the stationary solutions that have near equal values of both 
Hamiltonian and norm there is likely no trajectory in phase space passing 
close to them both. Thus, in order to have good mobility of narrow modes in 
discrete systems, we need \emph{not only} a small PN-barrier but also an 
exchange of stability between $1$-site and $2$-site solutions taking place 
\emph{in the vicinity}. We therefore also conjecture that the size of the 
oscillating background accompanying travelling modes~\cite{Gomez.04,Gomez.06} 
is not only related to the size of the PN-barrier but also to the difference 
in other conserved quantities (like $\mc{N}$) of the involved stationary modes 
measured at the points where their stability is changed. The lack of mobility 
near a zero of the energy difference between fundamental modes was observed 
also in~\cite{Maluckov.05}, but no further analysis was carried out.

Effects of a more delicate balance between linear and nonlinear coupling terms 
in Eq.~\eref{eq:xdnls} have also been discussed. These include the existence 
and stability of compact solutions, both discrete breathers and discrete 
vortex solitons. The latter have to our knowledge not been previously 
reported. Mathematically, these solutions are not robust with respect 
to parameter variation and require a balance of the coupling parameters in 
the equation for exact compactness, but near perfect localization persists in 
the neighborhood of the exact parameter values. An interesting question is if 
they can prevail also beyond the tight-binding approximation of 
Eq.~\eref{eq:xdnls}. The answer is likely that they are only exact solutions 
for the present model, but that they will correspond to modes of a higher 
degree of localization in more accurate models, i.e., they give an indication 
in which parameter regimes high localization can be achieved. As discussed in 
Sec.~\ref{sec:compact} the compact solutions do not represent a true 
compactification in the real system.

Additional interesting effects present also in 1D arise from a non-trivial 
dependence of the power flow in the lattice on the amplitude and phase 
difference of an excitation. Particularly, the current can become zero for a 
non-trivial phase twist which also in 2D leads to the existence of complex 
localized stationary modes with an open geometry, that may even be stable. 
Further, we have shown how the transversal flow of power in the array of 
waveguides can be controlled with great flexibility by excitation with plane 
waves, and in particular explicitly demonstrated how the transport 
\emph{direction} may be continuously tuned by \emph{amplitude} variation. 
This may have applications for power-coupling devices. However, the use of a 
2D array of the type studied here for multiport switching is discouraged due 
to the poor mobility of localized modes. For this purpose a 1D 
array~\cite{Oster.03} or an array exhibiting a saturable~\cite{Vicencio.06}
or quadratic~\cite{Susanto.07} (see discussion below) 
nonlinearity instead of the Kerr nonlinearity is better suited.

We end by commenting briefly on a number of works on related issues, that 
appeared in print after the original submission of our manuscript. 

In~\cite{Abdullaev.08}, Abdullaev et.al extended the Wannier-function 
approach of~\cite{Alfimov.02} for the continuous 1D NLS equation with periodic
linear potential to the case with periodic variation also for the nonlinearity 
coefficient (which is the case also considered in our work). Within the 
tight-binding approximation, they derived under quite general conditions a
lattice equation, which in a special case ($\sigma=1$ 
in~\cite{Abdullaev.08}) is equivalent to the 1D version of our 
Eq.~(\ref{eq:xdnls}) as derived in~\cite{Oster.03}. An important conclusion 
of~\cite{Abdullaev.08} is the crucial importance to include also the nonlinear 
coupling terms (corresponding to our $Q_4$ and $Q_5$), as they for 
specific choices of nonlinear interactions were shown to be comparable with, 
or even stronger than, the on-site nonlinearity. Thus, this supports the 
soundness of our 
approach to consider variation of the parameters $Q_4$ and $Q_5$ over 
a rather large range of values. 

Several works discussing properties of moving solitons and (vanishing) 
PN-barriers in 1D generalized DNLS models have appeared. 
Dmitriev, Khare et al.~\cite{Dmitriev.07,Khare.07} found analytically exact
stationary and moving solutions for some of the exceptional, translationally 
invariant discretizations of~\cite{Dmitriev.06,Pelinovsky.06,Kevrekidis.06}. 
Whether these are of relevance for any physically realizable system is, to 
our knowledge, still unclear. Pelinovsky et al.~\cite{Pelinovsky.07} 
developed a mathematical technique for analysis of persistence of traveling 
single-humped localized solutions from a certain limit, and found 
specifically that while travelling solutions terminated when continued from 
the integrable AL-limit of the Salerno model (corresponding 
to the development of a resonant tail~\cite{Gomez.04}), 
they generally persisted in 
the translationally invariant models. 
Oxtoby and Barashenkov~\cite{Oxtoby.07} used 
the method of asymptotics beyond all orders to evaluate the amplitude of 
radiation from a moving small-amplitude soliton in the saturable DNLS 
equation, and found it to be completely suppressed at certain 
`sliding velocities', where, similarly as in~\cite{Melvin.06}, it 
was interpreted as an embedded soliton. The properties of the travelling 
solitary waves in the saturable model were also analyzed numerically 
in more detail by Melvin et al.\ in~\cite{Melvin.08}. 

For the two-dimensional case, Susanto et al.~\cite{Susanto.07} studied the 
mobility of discrete solitons in a square lattice with \emph{quadratic} 
(second-harmonic-generating) nonlinearity. In this case, due to the absence 
of collapse instability in the continuum limit, good mobility of 
stable solutions was observed 
as long as the coupling constants were not too small 
(`quasicontinuum regime'). In this regime, mobility in arbitrary 
directions (not necessarily axial or diagonal) was observed, which is not 
surprising since discreteness effects are smoothened out for broad, 
continuum-like solitons. However, no non-trivial zeros of the PN barrier 
were found for this model, and consequently strongly localized solutions were 
reported to be immobile. Thus, in many aspects, the mobility properties 
of the 2D model with quadratic nonlinearity is similar to those of the 1D 
cubic on-site DNLS model.

Finally, a very recent preprint by Chong et al.~\cite{Chong.08} extended 
the analysis of the on-site cubic-quintic DNLS model 
of~\cite{Carretero.06} to higher dimensions. As concerns the two-dimensional 
mobility, the results were shown to be very similar to that of the saturable 
model~\cite{Vicencio.06}: enhanced mobility was found in regimes close to 
stability inversion and associated with appearance of 
asymmetric intermediate solutions and low PN barrier. Thus, as far as we are 
aware, the present work still provides the only known explicit example of 
a model where these properties do \emph{not} lead to an enhanced mobility.

\ack
M\"{O} would like to thank the Mathematics Department at Heriot-Watt 
University, Edinburgh, for its hospitality and especially J.C.\ Eilbeck for 
guidance, support and very useful discussions. This work was partly carried 
out under the HPC-EUROPA project (RII3-CT-2003-506079), with the support of 
the European Community - Research Infrastructure Action under the FP6 
"Structuring the European Research Area" Programme. Partial support from the 
Swedish Research Council is acknowledged.


\end{document}